%% file: moriond.tex
\documentclass[11pt]{article}
\usepackage{moriond,epsfig}
\usepackage{subfigure}
\def\Journal#1#2#3#4{{#1} {\bf #2}, #3 (#4)}

\def\NPB{{\em Nucl. Phys.} B}
\def\PLB{{\em Phys. Lett.}  B}

\def\EPJ{{\em Eur. Phys.J.} C}
\def\ZPC{{\em Z. Phys.} C}

\def\be{\begin{equation}}
\def\ee{\end{equation}}
\def\bea{\begin{eqnarray}}
\def\eea{\end{eqnarray}}

\include{macro}

\begin{document}
\vspace*{4cm}
\title{BOSON GAUGE COUPLINGS AT LEP}

\author{Paolo SPAGNOLO }

\address{Istituto Nazionale di Fisica Nucleare, Sezione di Pisa,\\
 Largo B. Pontecorvo, 3, 56100 PISA, Italy}

\maketitle\abstracts{ 
The review of the measurements of the gauge couplings in the boson sector at LEP is presented.
The measurements of the charged triple gauge couplings (cTGC) parameters from the four LEP experiments are combined 
and the results are in good agreement with the Standard Model predictions, proving the
non-abelian structure of the $ { \rm SU(2)_L \times U(1)_Y }$ gauge simmetry.
Different measurements of these parameters are reviewed and all possible fit methods discussed.
The current limits on the anomalous neutral triple gauge couplings (nTGC) and the quartic gauge coplings (QGC) 
are also presented.
}

\section{Introduction}

The LEP \eee collider has been the most important experimental environment to study the electroweak 
interactions and precisely test the Standard Model.
Since 1989 each of the four LEP experiments collected data corresponding to an integrated luminosity larger than 1000 pb$^-1$, of which 
about 200 pb$^-1$  at a centre-of-mass energy of the \Z mass peak (from 1989 to 1995).
After 1996 a second phase started, LEP2, with collected lumonosity of 750 pb$^-1$ per experiment 
and energy above the WW production threshold 
(centre-of-mass energy above 161 GeV), allowing to
produce  for the first time two real (on shell) massive bosons WW, ZZ, decaying in a four fermions final state.
These new channels opened at LEP2, allow the precise measurement of the mass of the W but are also 
sensitive to the trilinear gauge couplings (TGC) predicted in the Standard Model~\cite{teo}.
LEP ended collecting data in 2000 but only recently all the most important analyses in the W sector have been  
completed and the results of the four LEP experiments combined.
This proceeding is a review of the measurements of the boson gauge couplings at LEP.
All the results here presented have been published by ALEPH~\cite{al}, L3~\cite{l3}, OPAL~\cite{op} and DELPHI~\cite{de} collaborations. 

\section{Triple gauge boson couplings}

The triple gauge boson couplings (TGC) can be classified into charged couplings (cTGC)  
when the triple boson vertex is WW$\gamma$ or WWZ  and neutral couplings (nTGC) for the case of ZZZ, ZZ$\gamma$ or Z$\gamma$$\gamma$.
The first  are forseen in the Standard Model and are linked to the non-abelian structure of 
the $ { \rm SU(2)_L \times U(1)_Y }$ gauge simmetry. The neutral nTGC at the contrary, do not exist in the Standard Model at tree level.

\subsection{Charged triple gauge couplings}

In order to measure the charged triple gauge couplings (cTGC)
 at LEP is necessary to have a vertex with two W and a neutral boson (Z,$\gamma$). Within the Standard Model, this vertex 
can be achieved through the production of two massive W with the \WW process available for the first time at LEP2 or, 
with lower sensitivity, through the single W production with the process
\We and through the single photon production \singam.

\begin{figure}[Ht!]
\begin{center}
\mbox{
\subfigure[]{
\includegraphics[width=0.3\textwidth]{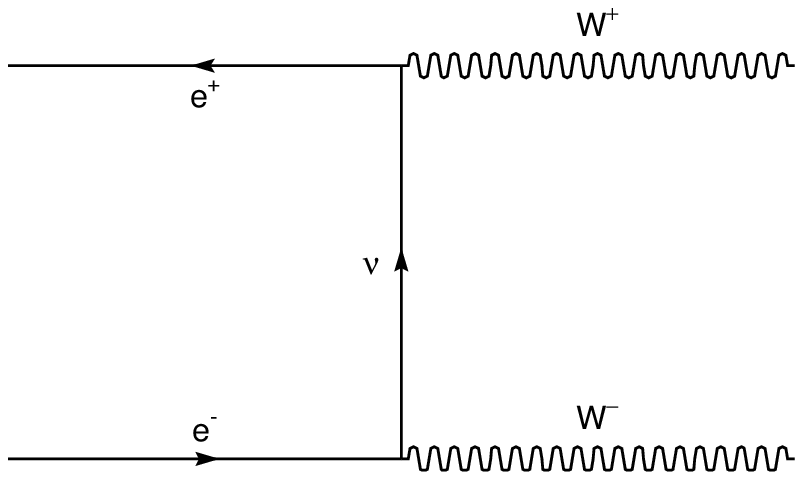}
}
\subfigure[]{
\includegraphics[width=0.3\textwidth]{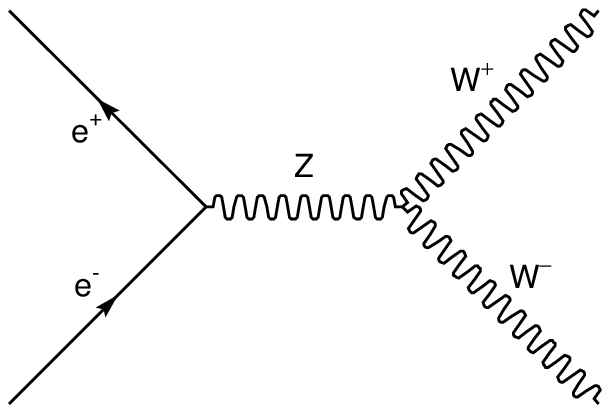}
}
\subfigure[]{
\includegraphics[width=0.3\textwidth]{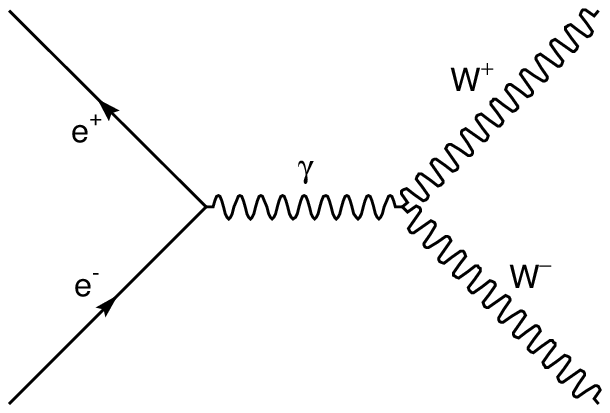}
}
}
\end{center}
\caption{Diagrams of the WW production at LEP2}
\label{ww} 
\end{figure}

The first proof of the existence of the charged TGC is in the measurement of the WW cross section.
Three diagrams are responsible for the WW production as shown in fig.~\ref{ww}:
the {\it t-channel} with the neutrino exchange (fig.~\ref{ww},a), the {\it s-channel} with a Z exchange  (fig.~\ref{ww},b)
and the {\it s-channel} with a virtual photon exchange  (fig.~\ref{ww},c). Both the {\it s-channel} diagrams have a TGC vertex. 

\begin{figure}[Htbp!] 
\begin{center}
\epsfig{figure=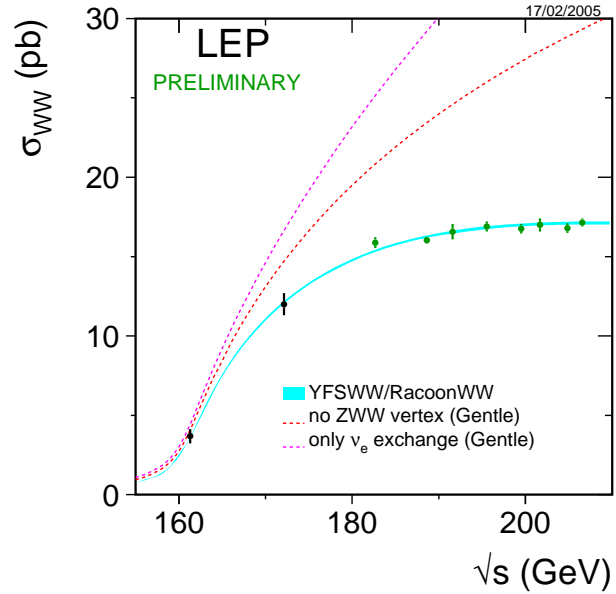,width=8cm}
\end{center}
\caption{WW cross section measurement at LEP}
\label{xsec} 
\end{figure}

All the three diagrams are needed to make the WW cross section converging to finite values at high energies.
If the $ { \rm SU(2)_L \times U(1)_Y }$ was an abelian gauge simmetry, the TGC verteces would be forbidden and 
the WW cross section would diverge when increasing the centre-of-mass energy. 
The LEP2 data fit precisely the Standard Model predictions and confirm the presence of the TGCs and the non-abelian structure of 
the $ { \rm SU(2)_L \times U(1)_Y }$ gauge simmetry, as can be viewed in fig.~\ref{xsec}.

The most general form for an effective charged TGC Lagrangian consistent with Lorentz invariance involves 14 complex couplings,
7 for the WW$\gamma$ and 7 for the WWZ vertex.
Most of these couplings are C- or P-violating while in the Standard Model C- and P-conservation are predicted in the TGCs sector.
Assuming C and P conservation the 14 complex couplings are reduced to 6 real couplings: 
$g_1^\gamma$, $g_1^Z$, $k_\gamma$, $k_Z$, $\lambda_\gamma$ and $\lambda_Z$.
In the Standard model $g_1^\gamma= 1$, $g_1^Z=1$, $k_\gamma=1$, $k_Z=1$, $\lambda_\gamma=0$ and $\lambda_Z=0$.
The couplings can be related to physical properties of the gauge bosons, like the electric dipole, quadrupole and magnetic moment. 
For instance the W anomalous magnetic moment can be written as 
\be
\mu_W = \frac{e}{2M_W}(1+k_\gamma + \lambda_\gamma )
\ee
The requirement of local  $ { \rm SU(2)_L \times U(1)_Y }$ gauge invariance introduce the further constraints
\bea
\Delta k_Z & = & -\Delta k_\gamma \tan^2 \theta_W + \Delta g_1^Z  \\
\lambda_\gamma & =  & \lambda_Z \nonumber  \\
g_1^\gamma &=& 1 \nonumber
\label{2}
\eea
with $\Delta$ indicating the deviation from the Standard Model predictions and $\theta_W$ the electroweak mixing angle,
leaving 3 independent real couplings~\cite{teo2}: $g_1^Z$, $k_\gamma$ and  $\lambda_\gamma$.

These couplings have been experimentally tested with the \WW sample collected at LEP2. The angular distribution of the \WW
 cross section are more sensitive to TGCs than the inclusive measurement. 
A precise study of the TGC parameters is achieved with the analysis of the differential WW cross sections respect to 5 angles:
\begin{itemize}
\item the angle $\theta_W$ between the $W^-$ and the initial $e^-$ in the WW rest frame
\item the polar and azimuthal angles of the fermion in the decay $W^- \to f \bar f  $ calculated in the rest frame of the $W^-$
\item the corresponding polar and azimuthal angles of the fermion in the decay of the $W^+$. 
\end{itemize}
All the possible WW decays, shown in tab.~\ref{BR} together with the branching ratios and the typical efficiencies,
are taken into account for this analysis. 

\begin{table}[t]
\caption{\it 
WW decay modes with relative BR, efficiencies.
}
\begin{center}
\begin{tabular}{|l|c|c|} \hline
Decay Mode  & BR & averaged $\varepsilon$ \\
\hline\hline
qqqq         & 45 \% & 85 \% \\
$\mu \nu$ qq   & 15 \% & 80 \% \\            
e $\nu$ qq     & 15 \% & 80 \% \\
$\tau \nu$ qq  & 15 \% & 60 \% \\
$\ell \nu \ell \nu$ & 10 \% & 65 \% \\
\hline
\end{tabular}
\end{center}
\label{BR}
\end{table}

In the semileptonic $W^+ W^- \to qq l \nu$ decays the charge of the lepton identifies without ambiguity the W$^-$ and 
the missing momentum provides information on the direction and the energy of the undetected neutrino. Since quark and anti-quark
 are not identified in the hadronic W decays, there are two ambiguous solutions that normally enter both with a weight 0.5 in the 
angular distributions used for the measurements. Ambiguities are more important in the fully hadronic and fully leptonic channels.
In the first case jets are paired using a likelihood fit to the W mass and the correct charge is extracted from an estimator.
In the second case the consistency between the W mass and the mass of the $l \nu$ system helps in reconstructing the neutrino momenta.
The quadratic nature of the constraints, however, always yield to a two-fold ambiguity: solutions obtained by flipping both neutrinos are 
equally valid. Again, both solutions enter in the experimental distribution with equal weight.

\begin{figure}[Htbp!] 
\begin{center}
\mbox{
\subfigure[]{
\includegraphics[width=0.27\textwidth]{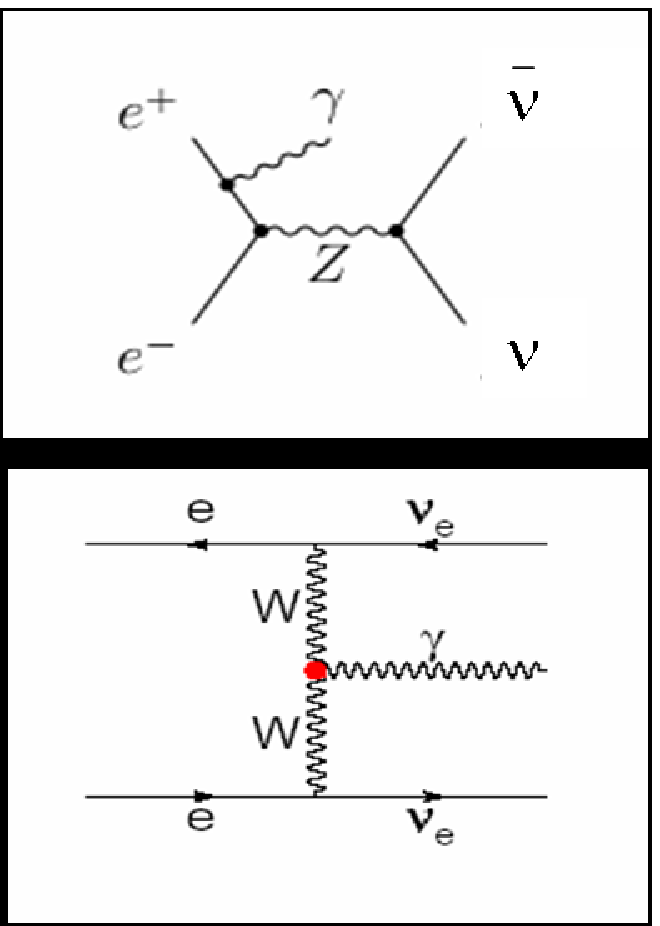}
}
\subfigure[]{
\includegraphics[width=0.4\textwidth]{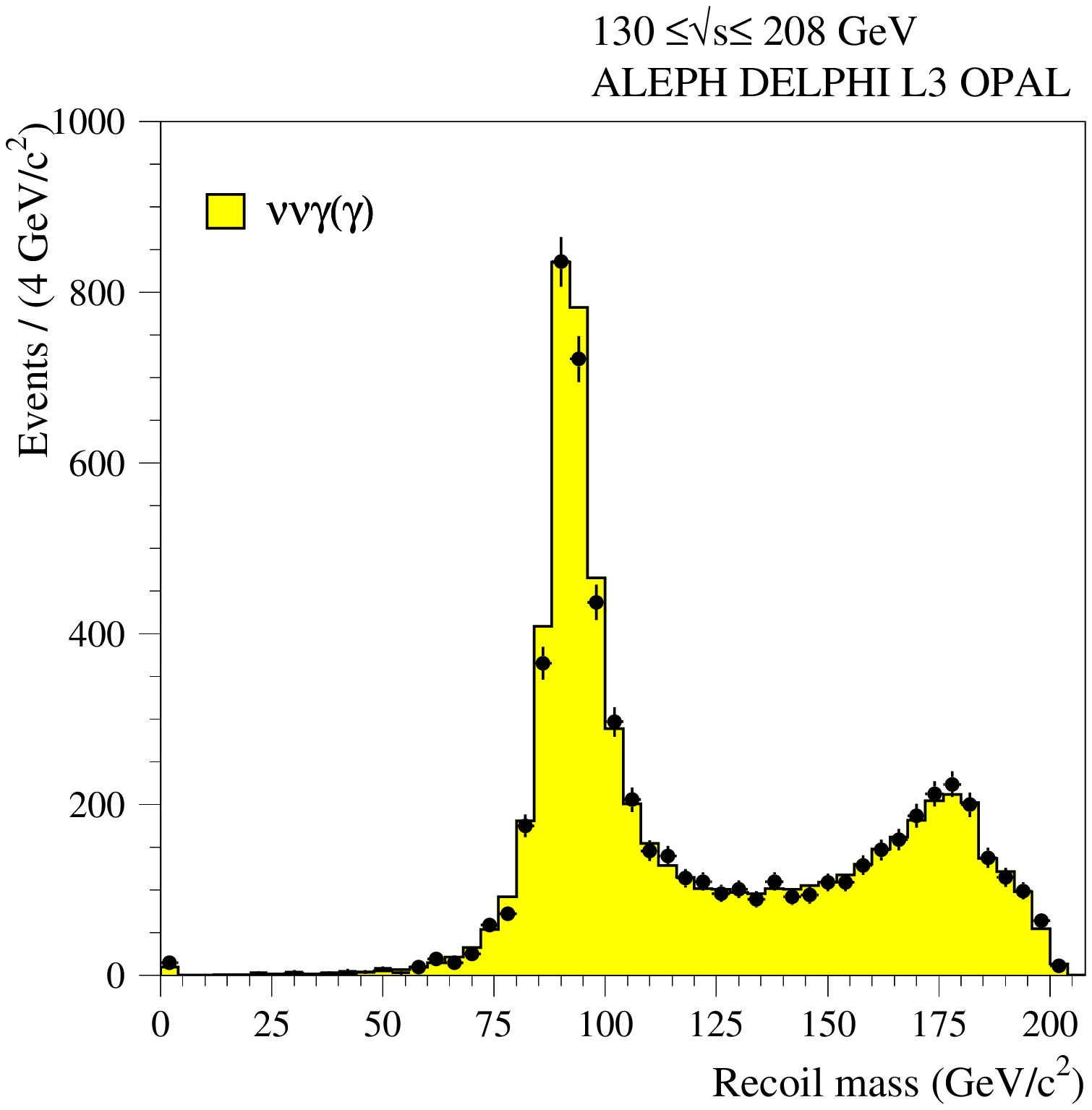}
}
}
\end{center}
\caption{Recoil mass in single photon events at LEP2}
\label{recoil} 
\end{figure}

The charged TGCs are measured by the LEP experiments using the also the  \We~ \\ and \singam events that are sensitive to the WW$\gamma$~ 
vertex. The \We events are reconstructed both in the leptonic \Wlnu  and in the hadronic \Wqq decay channels.
The leptonic channel selection requires a single good track identified as an electron or a muon with a missing energy and momentum.
The hadronic selection requires two hadronic jets and a missing energy an momentum with an angular distribution not compatible with a 
WW event. The \singam events are obtained through two different processes shown in fig.(~\ref{recoil}, a): 
the radiative return to the Z and the WW fusion. 
The two processes are well separated in the distribution of the recoil mass from the photon, in fig.(~\ref{recoil}, b).  
While the first kind of events allow an alternative way to measure the number of neutrino family, by counting the number of 
events under the Z mass peak in  fig.~\ref{recoil}, the WW fusion processes, above the the Z mass peak, are sensitive to the TGC.

\begin{figure}[Htbp!] 
\begin{center}
\epsfig{figure=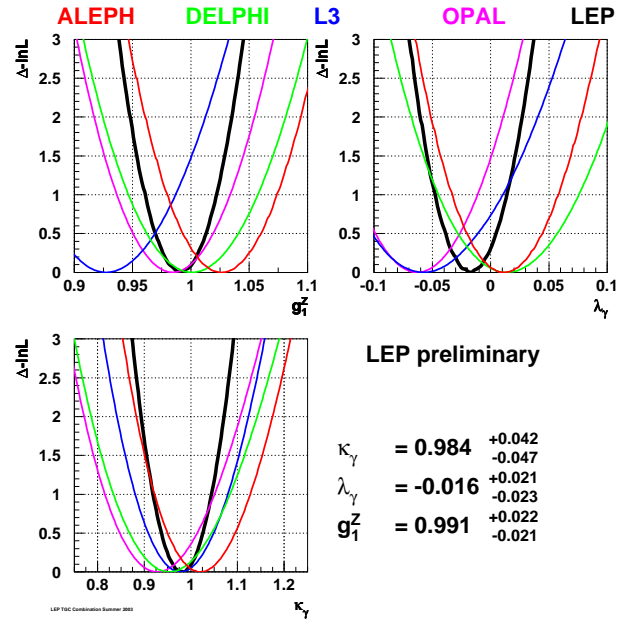,width=8cm}
\end{center}
\caption{Fit results on the triple gauge couplings.}
\label{lepfit} 
\end{figure}

Fits to the triple gauge couplings are performed with methods where only one parameter is allowed to vary and the other 
two are fixed to their Standard Model prediction to improve the precision of measurement.
The constraints obtained on the three triple gauge couplings from the combination of the LEP experiments~\cite{lep} 
are shown in fig.~\ref{lepfit}.

\begin{figure}[Htbp!] 
\begin{center}
\epsfig{figure=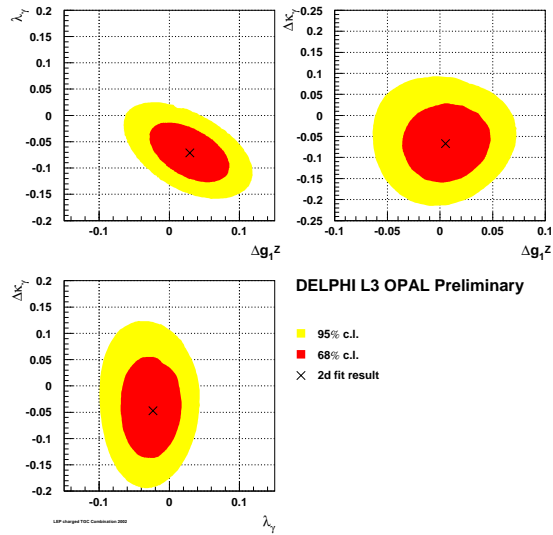,width=8cm}
\end{center}
\caption{Two dimension fit results on the triple gauge couplings.}
\label{2d} 
\end{figure}

To study the correlations in the measurements of the TGC couplings also fits in 3-dimension allowing the 3 parameter
free to vary  or in 2-dimension, constraining one parameter to its  Standard Model value and fit the other two, are implemented. 
The two dimension fit results are plotted in in fig.~\ref{2d}.

All the three cuplings are consistent with the Standard Model expectation, with a few percent precision.

ALEPH also performed a fit to all the 14 complex couplings, relaxing all the constraints on C- and P-conservation and 
$ { \rm SU(2)_L \times U(1)_Y }$ gauge invariance~\cite{al}. Out of all the 28 real parameters, one at the time is allowed to vary 
and the other are fixed to the Standard model predictions. The results of this test are:
\bea
Re(g_1^\gamma) &=& 1.123 \pm 0.091 \\
Re(g_1^Z)      &=& 1.066 \pm 0.073 \nonumber \\
Re(k_\gamma)   &=& 1.071 \pm 0.062 \nonumber \\
Re(k_Z)        &=& 1.065 \pm 0.061 \nonumber  
\eea 
All the other 24 parameters are consistent with zero  ($<$ 0.05 $\div $ 0.20 at  95 $\%$ C.L.), in perfect agreement with the Standard 
Model.

\subsection{Neutral triple gauge couplings}

Neutral TGC, as ZZZ, ZZ$\gamma$ or Z$\gamma$$\gamma$  do not exist in the Standard Model at tree level and are negligible 
at the LEP energies. The presence of anomalous neutral couplings would affect the \ZZ processes. The couplings describing the 
nTGC~\cite{teo3}
in the on shell ZZ production are 4, namely: $f_4^\gamma$, $f_5^\gamma$, $f_4^Z$ and $f_5^Z$, while 8 parameters are needed to describe
the on shell production of Z$\gamma$:  $h_1^\gamma$,  $h_2^\gamma$,  $h_3^\gamma$,  $h_4^\gamma$,  
$h_1^Z$,  $h_2^Z$,  $h_3^Z$,  $h_4^Z$.
Similarly to the WW case, the differential cross section of the ZZ events respect to the Z polar angle and 
the angular distribution distributions of the fermions from the Z decay are taken into account, to estimate the nTGC couplings.
Fit results are shown in tab.~\ref{ntgc}. 

\begin{table}[t]
\caption{\it 
LEP combined results of the fit to the nTGC parameters.
}
\begin{center}
\begin{tabular}{|l|c|} \hline
Parameter  & 95\% C.L. \\
\hline\hline
$h_1^\gamma$  & [-0.056, +0.055]  \\
$h_2^\gamma$  & [-0.045, +0.025]  \\
$h_3^\gamma$  & [-0.049, +0.008]  \\ 
$h_4^\gamma$  & [-0.002, +0.034]  \\ 
\hline\hline 
$h_1^Z$       & [-0.13,  +0.13]   \\
$h_2^Z$       & [-0.078, +0.071]  \\ 
$h_3^Z$       & [-0.20,  +0.07]  \\  
$h_4^Z$       &  [-0.05, +0.12]  \\
\hline\hline
$f_4^\gamma$  & [-0.17, +0.19]  \\
$f_5^\gamma$  & [-0.36, +0.40]  \\
$f_4^Z$       & [-0.31,  +0.28]  \\  
$f_5^Z$       &  [-0.36, +0.39]  \\
\hline
\end{tabular}
\end{center}
\label{ntgc}
\end{table}
All these couplings are compatible with zero therefore  
no departure from the cross sections predicted by the Standard Model for these processes is observed. 

\section{Quartic Gauge Couplings}

The Standard Model predicts the existence of verteces with four gauge bosons, like WW$\gamma \gamma$, WWZ$\gamma$,
even if their rates are negligible at LEP energies. On the other hand, as in the case of the TGCs, 
the completely neutral verteces with four bosons, like ZZ$\gamma \gamma$, are not allowed in the Standard Model.
The processes sensitive to the charged QGC are $e^+ e^- \to W^+ W^- \gamma$~,
with a final state identical to that of a real photon radiative correction to the \WW process;
and the WW fusion with two photon in the final state $e^+ e^- \to \bar\nu \nu \gamma \gamma$.
The neutral QGC could be obtained through a process like  $e^+ e^- \to \gamma\gamma Z $~.

The quartic couplings QGC are not dimensionless and are always referred to a parameter $\Lambda$ which has the dimension
of a mass and is commonly set to the W mass: $\Lambda = M_W$.  
Three parameters are used to describe the charged QGC, $a^W_0$,  $a^W_c$ and  $a^W_n$ 
while only two parameters are necessary for the neutral QGC:  $a^Z_0$ and  $a^Z_c$.
The constraints on  these parameters, combining the four LEP experiments fit results, are showed intab.~\ref{qgc}. 

\begin{table}[t]
\caption{\it 
LEP combined results of the fit to the QGC parameters.
}
\begin{center}
\begin{tabular}{|l|c|} \hline
Parameter  & 95\% C.L. \\
\hline\hline
$a^W_0/\Lambda^2$  & [-0.018, +0.018]  \\
$a^W_c/\Lambda^2$  & [-0.056, +0.012]  \\
$a^W_n/\Lambda^2$ & [-0.139, +0.114]  \\ 
\hline\hline 
$a^Z_0/\Lambda^2$ &  [-0.008, +0.021]  \\ 
$a^Z_c/\Lambda^2$ &  [-0.029, +0.039]  \\
\hline
\end{tabular}
\end{center}
\label{qgc}
\end{table} 

In both neutral and charged QGC no evidence of these anomalous couplings and therefore no deviation 
from the Standard Model has been observed.

\section{Conclusions}
The trilinear couplings TGC have been measured at LEP2 with a precision of few percent.
The proof of the non-abelian stucture of the $ { \rm SU(2)_L \times U(1)_Y }$ gauge simmetry
is obtained for the first time and this is one of the most important results of LEP2.
No deviation from the Standard Model prediction is observed.
The search of possible anomalous couplings 
for the trilinear neutral gauge couplings and the quartic gauge couplings is also presented. 
All the combined results are in good agreement with the Standard Model and no anomalous coupling
has found. 

\section*{Acknowledgments}

I would like to thank  Paolo Azzurri, Timothy Barklow, Stephane Jezequel, Roberto Tenchini and Andrea Venturi for their help 
in preparing this talk and the LEP Electroweak Working Group for the combination results and the plots here presented.

\end{document}

%% file: macro.tex
\def \rightdownarrow
{\kern.1em
\rule[.5ex]{.15mm}{3ex}
{\mbox{$\kern-0.1em{\longrightarrow}$}}
}

\def\lessim{\mathrel {\vcenter {\baselineskip 0pt \kern 0pt
\hbox{$<$} \kern 0pt \hbox{$\sim$} }}}
\def\gessim{\mathrel {\vcenter {\baselineskip 0pt \kern 0pt
\hbox{$>$} \kern 0pt \hbox{$\sim$} }}}

\newcommand{\We}{\mbox{$e^+ e^- \to W^{\mp} e^{\pm} \nu(\bar{\nu}) $~}}
\newcommand{\singam}{\mbox{$e^+ e^- \to \gamma  \bar{\nu_e}  \nu_e  $~}}
\newcommand{\ZZ}{\mbox{$e^+ e^- \to Z(\gamma^*) Z(\gamma^*) $~}}
\newcommand{\WW}{\mbox{$e^+ e^- \to W^+ W^- $~}}

\newcommand{\eee}{\mbox{$ e^+ e^- $}}
\newcommand{\Z}{\mbox{$Z\ $}}

\newcommand{\Wqq}{$W\to q\bar q \ $}
\newcommand{\Wlnu}{$W\to l\nu \ $}